\documentclass[a4paper,10pt]{revtex4-2}
\pdfoutput=1
\usepackage{comment}
\usepackage{color}
\usepackage{tabularx, booktabs}
\usepackage{multirow}
\usepackage{verbatim}
\usepackage{appendix}
\usepackage[labelformat=simple]{subcaption}

\DeclareCaptionLabelFormat{subcaptionlabel}{\normalfont(\textbf{#2}\normalfont)}
\usepackage{pgfplots}
\usepackage{lineno}
\pgfplotsset{compat=newest,
            every axis/.append style={
                    tick label style={font=\small},
                    legend style={font=\small}
                    }}
\usepackage{tikz}
\usetikzlibrary{quantikz} 
\usetikzlibrary{calc} 
\usetikzlibrary{external} 

\usepackage{braket} 
\usepackage[normalem]{ulem}

\newcommand{\bluebkg}{blue!5}
\newcommand{\bluebox}{blue!20}
\newcommand{\redbkg}{red!5}
\newcommand{\redbox}{red!20}

\pgfmathdeclarefunction{poisson}{1}{%
  \pgfmathparse{(#1^x)*exp(-#1)/(x!)}%
  }

\pgfplotstableread[col sep=comma]{./data/fidelity.csv}\fidelitydata
\pgfplotstableread[col sep=comma]{./data/occupation_prob.csv}\occupationprobdata
\pgfplotstableread[col sep=comma]{./data/ansatz_fidelity_depths.csv}\ansatzfidelitydata
\pgfplotstableread[col sep=comma]{./data/depth_iters.csv}\depthitersdata

\begin{document}
\title{Digital Quantum Simulation and Circuit Learning for the Generation of Coherent States}

\author{Ruilin Liu 
 $^{1}$, Sebasti\'{a}n V. Romero
 $^{2}$, Izaskun Oregi $^{2,3}$, Eneko Osaba $^{2}$, Esther Villar-Rodriguez $^{2}$ {and Yue Ban $^{2,}$}}
 \affiliation{ $^{1)}$School of Materials Science and Engineering, Shanghai University, Shanghai 200444, China}
 \affiliation{ $^{2)}$TECNALIA, Basque Research and Technology Alliance (BRTA), 48160 Derio, Spain}
 \affiliation{ $^{3)}$Faculty of New Interactive Technologies, Universidad EUNEIZ, 01013 Vitoria-Gasteiz, Araba, Spain}

\begin{abstract}
Coherent states, known as displaced vacuum states, play an important role in quantum information processing, quantum machine learning,
 and quantum optics. In this article, two ways to digitally prepare coherent states in quantum circuits are introduced. First, we construct the displacement operator by decomposing it into Pauli matrices via ladder operators, i.e., creation and annihilation operators. The high fidelity of the digitally generated coherent states is verified compared with the Poissonian distribution in Fock space. Secondly, by using Variational Quantum Algorithms, we choose different ansatzes to generate coherent states. The quantum resources---such as numbers of quantum gates, layers and iterations---are analyzed for quantum circuit learning. The simulation results show that quantum circuit learning can provide high fidelity on learning coherent states by choosing appropriate ansatzes.

\end{abstract}
\maketitle

\section{Introduction}
The exponentially increasing scaling of degrees of freedom poses challenges in the computation for quantum chemistry \cite{Quantum-Chemistry-review}. Recent developments in quantum computing present new routes for the exploration of quantum chemistry by taking advantage of quantum resources and manipulating the states of matter. In the Noisy Intermediate-Scale Quantum (NISQ) era \cite{NISQ}, seeking appropriate algorithms \cite{NISQ-algorithms} is crucial for quantum dynamics digital simulation, which is one of the most promising application{s}.

As the first step of any quantum algorithm, the preparation of an initial state directly decides its success. Furthermore, in the 
Hamiltonian simulation,
 an efficient and accurate implementation of $U = e^{-iHt}$ is another crucial task. 
 While the Hamiltonian of a molecule can be expressed easily in terms of its first quantization in real space \cite{simulation-realspace} and simulate the kinetic part

using the Quantum Fourier Transformation \cite{simulation-QFT}, the basis set for expressing the quantum states is smaller and more explicit in the second quantization \cite{simulation-second-quantization}, since a Fock state can be easily represented in the computational basis.

Analyzing and improving asymptotic scaling of resources on a quantum computer have
 been widely studied during the last decade \cite{NISQ-algorithms,Q-algorithms1, Q-algorithms2, Q-algorithms3}. In the NISQ era, low-depth circuits and reduced number of quantum gates are required to execute quantum algorithms in the presence of limited coherence time. Variational Quantum algorithms (VQA) \cite{VQA1,VQA2,VQA3}, which are hybrid quantum-classical methods that take advantage from quantum and classical computation, have demonstrated to be resource-efficient strategies. Among them, Variational Quantum Eigensolvers (VQE) \cite{VQE1, VQE2} can prepare the initial state and estimate the ground-state energy in a flexible and efficient way by choosing a suitable ansatz \cite{Hardware-efficient-ansatz,UCC-ansatz,adaptive-VQE} in a parameterized quantum circuit (PQC). In addition, optimization in PQCs aims at achieving high fidelity of digital simulation by mitigating quantum errors \cite{error-mitigation-simulation}. To this end, gradient-descent based quantum circuit learning (QCL) \cite{QCL} can be used to optimize the control function via finding out appropriate variational parameters.

Coherent states are a very special set of states forming the basis of
continuous variables in quantum information \cite{CV-QI1, CV-QI2}, being useful in a wide variety of applications---for instance,
 to represent thermal and Schr\"{o}dinger cat states, in the Mach--Zehnder interferometers configuration \cite{MZ-interferometer1,MZ-interferometer2}, in quantum metrology \cite{Q-metrology1,Q-metrology2}, in quantum cryptography \cite{Q-cryptography}, {in quantum machine learning \cite{machine-learning1,machine-learning2}}, among others. 
 Simulation on coherent states in quantum circuits provides a fundamental and digital alternative to design and optimize quantum phenomena, quantum dynamics in gate-based quantum computers.
Since the scalability of quantum algorithms is still limited by the complexity of quantum circuits, high-fidelity and efficient ways for digital quantum simulation of coherent states are indispensable for quantum technologies. 
In this paper, we propose a new method of digitized quantum simulation to simulate and optimize the generation of coherent states. We first use the creation and annihilation operators of the quantum harmonic oscillator to construct the displacement operator in the basis of a Fock state.
 By mapping the displacement operator acting on the vacuum state in the circuit, we can obtain
  coherent states with high fidelity.  At the same time, to reduce the consumption of quantum resources in the digitized quantum simulation process, we propose VQA by combining the quantum circuit with a \mbox{classical optimizer}.\par
The paper is organized as follows. In Section \ref{coherentstate}, we introduce the coherent state and its digital simulation in a quantum circuit by trotterizing the displacement operator, where its decomposition with $N$ qubits is generalized via the creation and annihilation operators. In Section \ref{Variationalquantumalgorithm}, 
the 
Hardware Efficient Ansatz and Checkerboard Ansatz \cite{Checker-board-ansatz}
are used to learn the coherent state. The quantum resources such as the number of quantum gates and iteration times are compared in different schemes. Finally, we give the Conclusion in Section \ref{conclusion}.

\section{Coherent State and Its Digital Simulation}
\label{coherentstate}
First consider a quantum harmonic oscillator with the time-independent frequency $\omega$, whose Hamiltonian can be written as ($m= \hbar  = 1$ in dimensionless units{, where $m$ is mass and $\hbar$ is the 
reduced Planck constant}):
\begin{equation}
H=\omega\left(\hat{a}^\dagger\hat{a}+\frac{1}{2}\right) \label{H}.
\end{equation}

In the basis of a Fock space $\{\ket{n}\}$, the creation operator $\hat{a}^\dagger$ and the annihilation operator $\hat{a}$ are written respectively as

\begin{equation}
\hat{a}^\dagger =
\left  [
\begin{matrix}
0 & 0 & 0 & \cdots & 0 &\cdots \\
\sqrt{1} & 0 & 0 & \cdots & 0 &\cdots \\
0 & \sqrt{2} & 0 & \cdots & 0 &\cdots \\
\vdots & \vdots & \ddots & \ddots & \cdots & \cdots \\
0 & 0 & \cdots & \sqrt{n}  & 0 & \cdots\\
\vdots & \vdots & \vdots & \vdots&\ddots&\ddots \\
\end{matrix}
\right ], \qquad 
\hat{a}= 
\left  [
\begin{matrix}
0 & \sqrt{1} & 0  & \cdots & 0 & \cdots\\
0 & 0 & \sqrt{2}  & \cdots & 0 & \cdots\\
0 & 0 & 0 & \ddots &\vdots &\cdots\\
\vdots & \vdots & \vdots & \ddots & \sqrt{n} & \cdots \\
0 & 0 & 0 &\cdots &0 &\ddots \\
\vdots & \vdots & \vdots &\vdots &\vdots &\ddots 
\end{matrix}
\right ].
\end{equation}

Using the displacement operator $\hat{D}(\alpha) = e^{\alpha\hat{a}^\dagger - \alpha^*\hat{a}}$ to act on the vacuum state $\ket{0}$, we can generate the coherent state $\ket{\alpha}=\hat{D}(\alpha)\ket{0}$, which is also the eigenstate of the
annihilation operator $\hat{a}$, satisfying
 $\hat{a}\ket{\alpha} = \alpha\ket{\alpha}$ with the arbitrary complex number $\alpha$. As in Fock space $N$ qubits map $n=2^N$ number of states, we directly
express the coherent states in the basis of Fock states $\{\ket{n}\}$.
It is feasible to simulate the generation of coherent states digitally in a quantum circuit, since the displacement operator $\hat{D}(\alpha)$ is a unitary operator.

Without loss of generality, we consider $\alpha$ to be a complex number $a+bi$, where $a$ and $b$ are real numbers and $i$ is the imaginary unit, so that the displacement operator is written as 
\begin{align}
\hat{D}(\alpha)
=e^{ (a+bi)\hat{a}^\dagger - (a-bi)\hat{a} }.
\label{2 qubit complex }
\end{align}

Here, we use the Hermitian matrix $Z_1 = i(a^\dagger -a)$ and the symmetric one $Z_2 = -(\hat{a}+\hat{a}^\dagger)$, where 
\begin{equation}\label{eq:z1_z2}
 Z_k = i^k
    \begin{bmatrix}
        0 & (-1)^k\sqrt{1} & 0 & \dots & 0 \\
        \sqrt{1} & 0 & (-1)^k\sqrt{2} & \dots & 0 \\
        0 & \sqrt{2} & \ddots & \ddots & \vdots \\
        \vdots & \vdots & \ddots & \vdots & (-1)^k\sqrt{n+1} \\
        0 & 0 & \dots & \sqrt{n+1} & 0
    \end{bmatrix},\quad k\in\{1,2\},
\end{equation}
is truncated from the subspace-embracing $N$ qubits. 
For $N$ qubits, we need $N2^{N-1}$ Pauli strings for decomposing $Z_1$ and $Z_2$ by using  $\smash[b]{\sigma_{\{0,1\}}^D\in\{I, \sigma_z\}}$ (diagonal Pauli matrix) and $\smash[b]{\sigma_{\{0,1\}}^S\in\{\sigma_x,\sigma_y\}}$ (skew-diagonal Pauli matrix). 
To express the formula in a more compact way, we can rewrite $Z_1= \sum_{l=1}^{N2^{N-1}}c_lA_1(l)$ and $Z_2= -\sum_{l=1}^{N2^{N-1}}c_l A_2(l)$, where $c_l$ and $A_{1,2}(l)$ are suitable definitions of each of the $N2^{N-1}$ constants and strings, respectively (see \mbox{Appendix \ref{APPENDIX-A}} for more details).
By using the first-order Suzuki--Trotter formula \cite{Suzuki,Trotter}, we can decompose the displacement operator into $M$ number of Trotter steps, where
\begin{equation}
\hat{D}(\alpha) \approx \left(e^{-iaZ_1/M}e^{-ibZ_2/M}\right)^M 
  \\ = \left[\left(\prod_{l=1}^{N2^{N-1}}e^{-iac_lA_1(l)/M}\right)\left(\prod_{l=1}^{N2^{N-1}}e^{ibc_lA_2(l)/M}\right)\right]^M.
\label{D-trotter}
\end{equation} 

In fact, $\hat{D}(a)$ and $\hat{D}(ib)$ in Equation \ref{2 qubit complex }, representing the real and the imaginary displacements of the coherent state, respectively, are implemented in quantum circuits in completely different ways due to their distinct decomposition into tensor products of the matrices from $\sigma^D$ and $\sigma^S$. 
We use Qiskit \cite{QISKIT} to implement the quantum circuit. Initialized in the vacuum states, i.e., $\ket{0}$ in all qubits, the quantum circuit generates coherent states after the implementation of the displacement operator (Equation \ref{D-trotter}) into a finite number of segmentations. 
For instance, a two-qubit system ($N=2$) with four Fock states 
expressed in the computational basis, leads to the following real and imaginary parts for the \mbox{displacement operator:}

\begin{align}
&\hat{D}(a) = e^{-i a Z_1} = e^{-ia [\frac{1+\sqrt{3}}{2}\sigma_y^1+\frac{1-\sqrt{3}}{2}(\sigma_z^0 \bigotimes  \sigma_y^1) + \frac{\sqrt{2}}{2}(\sigma_y^0 \bigotimes  \sigma_x^1-\sigma_x^0\bigotimes \sigma_y^1)]},\\
&\hat{D}(ib) = e^{-i b Z_2} 
= e^{ib [\frac{1+\sqrt{3}}{2}\sigma_x^1+\frac{1-\sqrt{3}}{2}(\sigma_z^0\bigotimes\sigma_x^1) + \frac{\sqrt{2}}{2}(\sigma_x^0 \bigotimes \sigma_x^1-\sigma_y^0\bigotimes \sigma_y^1)]}. 
\label{2 qubit d imaginary and real}
\end{align} 

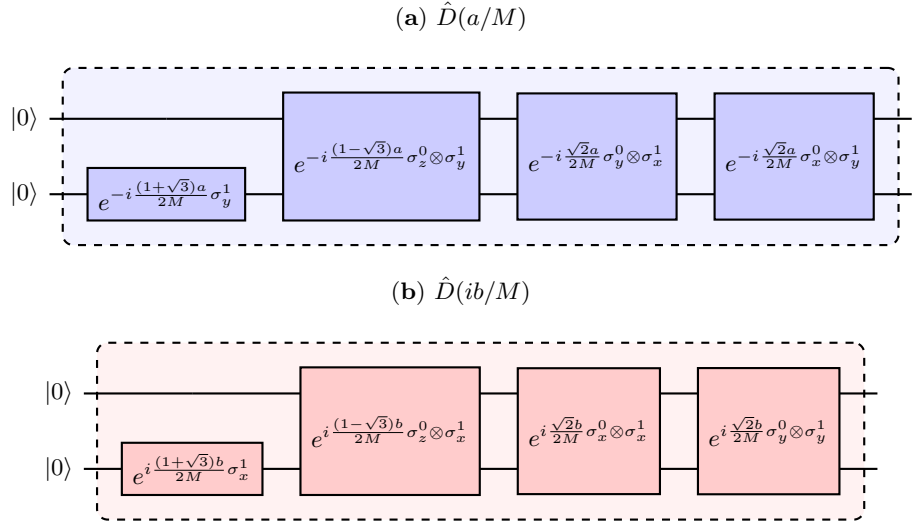
\begin{figure}[!htbp]
 \centering
    \begin{subfigure}[t]{\textwidth}\centering
    \caption{$\hat{D}(a/M)$}
    \begin{quantikz}[row sep={1cm,between origins}]
        \lstick{$\ket{0}$} & \qw\gategroup[wires=2,steps=4,style={dashed, inner sep=6pt, rounded corners, fill=\bluebkg}, background]{} & \gate[style={fill=\bluebox}, wires=2]{e^{-i\frac{(1-\sqrt{3})a}{2M}\sigma_z^0\otimes\sigma_y^1}} & \gate[style={fill=\bluebox}, wires=2]{e^{-i\frac{\sqrt{2}a}{2M}\sigma_y^0\otimes\sigma_x^1}} & \gate[style={fill=\bluebox}, wires=2]{e^{-i\frac{\sqrt{2}a}{2M}\sigma_x^0\otimes\sigma_y^1}} & \qw \\
        \lstick{$\ket{0}$} & \gate[style={fill=\bluebox}]{e^{-i\frac{(1+\sqrt{3})a}{2M}\sigma_y^1}} & & & & \qw
    \end{quantikz}
    \end{subfigure}\\[.3cm]%
    \begin{subfigure}[t]{\textwidth}\centering
    \caption{$\hat{D}(ib/M)$}
    \begin{quantikz}[row sep={1cm,between origins}]
        \lstick{$\ket{0}$} & \qw\gategroup[wires=2,steps=4,style={dashed, inner sep=6pt, rounded corners, fill=\redbkg}, background]{} & \gate[style={fill=\redbox}, wires=2]{e^{i\frac{(1-\sqrt{3})b}{2M}\sigma_z^0\otimes\sigma_x^1}} & \gate[style={fill=\redbox}, wires=2]{e^{i\frac{\sqrt{2}b}{2M}\sigma_x^0\otimes\sigma_x^1}} & \gate[style={fill=\redbox}, wires=2]{e^{i\frac{\sqrt{2}b}{2M}\sigma_y^0\otimes\sigma_y^1}} & \qw \\
        \lstick{$\ket{0}$} & \gate[style={fill=\redbox}]{e^{i\frac{(1+\sqrt{3})b}{2M}\sigma_x^1}} & & & & \qw
    \end{quantikz}
    \end{subfigure}%
\caption{Diagram of a single step for a two-qubit circuit implementation of the displacement operator $\hat{D}$ with $M$ Trotter steps. The displacement is (\textbf{a}) a real number $a$ and (\textbf{b}) an imaginary value $ib$.}
\label{displacement_operator}
\end{figure}

For example, the digital implementation of the displacement operator with real and imaginary displacements $\hat{D}(a)$ and $\hat{D}(ib)$ in a single Trotter step, where the total Trotter number is $M$, expressed by two qubits is shown in Figure \ref{displacement_operator}.  

To analyze the accuracy of the coherent state we derive, we define its fidelity

\begin{equation}
    F = |\langle\psi_{\text{f}}|\psi_{\text{tar}}\rangle|^2
\end{equation}
where $\ket{\psi_{\text{f}}}$ is the final state prepared by the circuit and $\ket{\psi_{\text{tar}}}$ is the target state  which can be calculated in Fock space as

\begin{equation}
    \ket{\psi_{\text{tar}}} = \ket{\alpha} =\exp\left(-\frac{1}{2}|\alpha|^2\right)\sum_{k=0}^\infty \frac{\alpha^k}{\sqrt{k!}}\ket{k}.
    \label{target-state}
\end{equation}

Such a target has the probability of the $m$-th Fock state in the analytical form 
\begin{equation}
    P_m =  |\langle m|\psi_{\text{tar}}\rangle|^2 = e^{-|\alpha|^2} \frac{|\alpha|^{2m}}{m!} = e^{-\langle n \rangle} \frac{\langle n\rangle^m}{m!},
    \label{probability}
\end{equation}
which returns a Poissonian distribution centered at $\langle n\rangle$, with $\langle n \rangle = \langle \hat{a}^\dagger\hat{a} \rangle = |\alpha|^2$.

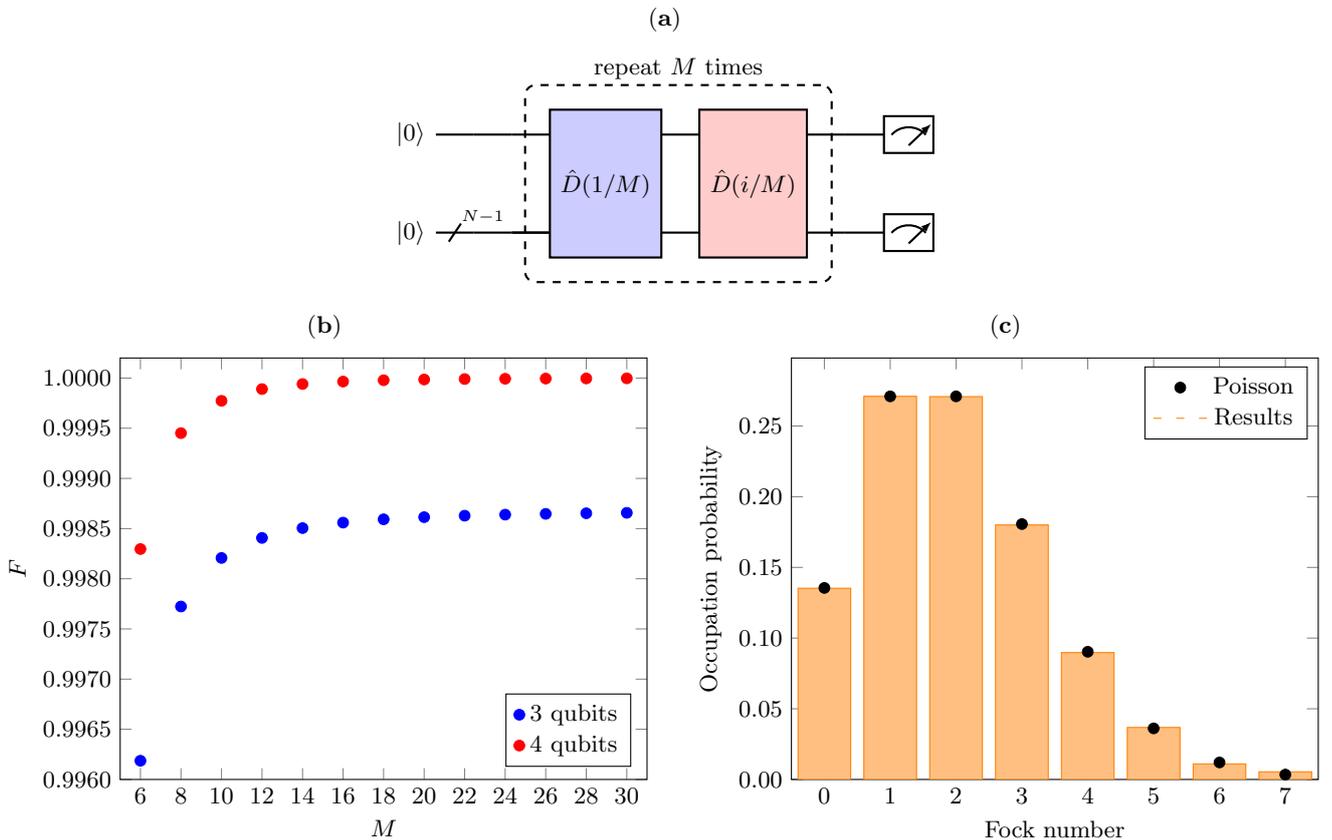
\begin{figure}[!t]
    \centering
    \begin{subfigure}[]{\textwidth}\centering
    \caption{}
    \label{fig:prep_coherent_state}
    \begin{quantikz}[row sep={1cm,between origins}]
        \lstick{$\ket{0}$} & \qw & \qw & \gate[style={fill=\bluebox}, wires=2]{\hat{D}(1/M)}\gategroup[wires=2,steps=2,style={dashed, inner sep=6pt, rounded corners}]{repeat $M$ times} & \gate[style={fill=\redbox}, wires=2]{\hat{D}(i/M)} & \qw & \meter{} \\ [.3cm]
        \lstick{$\ket{0}$} & \qwbundle{N-1} & \qw & \qw & & \qw & \meter{}
    \end{quantikz}
    \end{subfigure}\\[.3cm]%
    \begin{subfigure}[t]{0.5\textwidth}\centering
    \caption{}
    \label{fig:fidelities}
    \begin{tikzpicture}
        \begin{axis}[
            height=0.8\textwidth,
            width=0.95\textwidth,
            domain = 6:30,
            xlabel={$M$},
            ylabel={$F$},
            ytick distance=0.0005,
            xtick distance=2,
            y tick label style={
                /pgf/number format/.cd,
                fixed,
                fixed zerofill,
                precision=4,
                /tikz/.cd
            },
            ymax=1.0002, ymin=0.996,
            xmin=5, xmax=31,
            legend pos=south east,
        ]
        \addplot[blue, only marks] table[x=m, y=f_qub3]{\fidelitydata};\addlegendentry{3 qubits}
        \addplot[red, only marks] table[x=m, y=f_qub4]{\fidelitydata};\addlegendentry{4 qubits}
        \end{axis}
  \end{tikzpicture}
    \end{subfigure}%
    \begin{subfigure}[t]{0.5\textwidth}\centering
    \caption{}
    \label{fig:poisson}
    \begin{tikzpicture}
        \begin{axis}[
            height=0.8\textwidth,
            width=0.95\textwidth,
            xtick={0,1,...,7},
            domain = 0:7,
            samples = 8,
            xlabel={Fock number},
            ylabel={Occupation probability},
            ytick distance=0.05,
            y tick label style={
                /pgf/number format/.cd,
                fixed,
                fixed zerofill,
                precision=2,
                /tikz/.cd
            },
            ymin=0,
            xmin=-0.5, xmax=7.5,
        ]
        \addplot[ycomb, only marks] {1.001097923*poisson(2)};\addlegendentry{Poisson}
        \addplot[ybar, bar width=0.8, orange, fill=orange!50] table[x=fock_n, y=prob]{\occupationprobdata};\addlegendentry{Results}
        \end{axis}
  \end{tikzpicture}
    \end{subfigure}%
\caption{Preparation of the coherent state $\ket{\alpha} = |{1+i}\rangle$ in a quantum circuit. (\textbf{a}) Circuit implementation where the blue and the red blocks represent the parts with real and imaginary displacements, respectively. (\textbf{b}) The dependence of the fidelity of the coherent state $|{1+i}\rangle$ prepared by $3$ and \mbox{$4$ qubits} on the number of Trotter steps $M$. (\textbf{c}) Fock distribution of the coherent state ${\ket{1+i}}$, simulated by \mbox{$3$ qubits} and $M=20$ Trotter
 steps. The height of the yellow bar indicates the probability of finding particle in the $n$-th Fock state. The height in each Fock number coincides well with the Poissonian distribution (black-dotted) illustrated by $P_m$ (Equation \ref{probability}).}
\label{different_trotter_step}
\end{figure}

We benchmark our method by digitally generating the coherent state $\ket{\alpha}=|{1+i\rangle}$. Note that the coherent state is in bold in order to distinguish it from the Fock state $\ket{n}$. The circuit implementation is expressed in~Figure \ref{different_trotter_step}a, where the blue (real displacement part) and red (imaginary displacement part) blocks act alternatively. Depending on the magnitude of $|\alpha|$ and meanwhile aiming at achieving the desired fidelity, one needs to choose the appropriate number of qubits to derive the coherent state with a desired fidelity in the circuit. Using $4$ and $3$ qubits gives
 rise to the fidelities over $F = 0.9999$ with Trotter steps $M\geq14$ and $F=0.9986$ with $M\geq20$, respectively, as shown in~Figure \ref{different_trotter_step}b.
In~{Figure \ref{different_trotter_step}c}, we plot the distribution of the coherent state $\ket{\alpha}={\ket{1+i}}$ prepared by digitally implementing the displacement operator $\hat{D}$ with $3$ qubits and $M=20$ Trotter steps. It coincides well with the Poissonian probability distribution, indicated from Equation (\ref{probability}), where $\langle n \rangle = \langle \hat{a}^\dagger\hat{a} \rangle = |\alpha|^2 = 2$. The analysis on the accuracy of the coherent state expressed in a truncated space with $2^N$ states, where $N$ is the qubit number, can be found in Appendix \ref{APPENDIX-B}.

\section{Coherent State Generation by Variational Quantum Algorithm}
\label{Variationalquantumalgorithm}
Gradient-based quantum circuit learning, a kind of VQA for supervised learning, aims at achieving the target state by lowering the cost function. In this section, we use a variational quantum circuit combined with classical optimizers to prepare coherent states.
Inputting the ground state as the initial state into the circuit, processing the evolution in the form of three alternatives, we obtain the results at the output $\ket{\psi_\text{f}}$. To approach the target state $\ket{\psi_{\text{tar}}}$ (Equation \ref{target-state}) and by using gradient descent method based on the cost function
\begin{equation}
    C =  1 - |\langle\psi_\text{f}|\psi_{\text{tar}}\rangle|^2,
\label{cost-function}
\end{equation}
where $\ket{\psi_\text{f}}$ and $\ket{\psi_\text{tar}}$ are the final state and target state as defined above, we find the optimized variational parameters in the ansatz. Such VQA is implemented in the Qiskit package where the optimizer is SLSQP from SciPy~\cite{SLSQP}.

\begin{figure}[!htbp]
    \centering
    \begin{subfigure}[t]{\textwidth}\centering
        \caption{Scheme a}
        \label{fig:scheme_a}
        \begin{quantikz}[column sep={0.3cm}, row sep=0.1cm]
            & \gate{R_x(\theta_1)}\gategroup[wires=4,steps=7,style={dashed, inner sep=3pt, rounded corners}]{} & \gate{R_z(\theta_2)} & \gate{R_x(\theta_3)} & \ctrl{1} & \qw & \qw & \gate{R_y(\theta_{16})} & \qw \\
            & \gate{R_x(\theta_4)} & \gate{R_z(\theta_5)} & \gate{R_x(\theta_6)} & \gate{R_y(\theta_{13})} & \ctrl{1} & \qw & \qw & \qw \\
            & \gate{R_x(\theta_7)} & \gate{R_z(\theta_8)} & \gate{R_x(\theta_9)} & \qw & \gate{R_y(\theta_{14})} & \ctrl{1} & \qw & \qw \\
            & \gate{R_x(\theta_{10})} & \gate{R_z(\theta_{11})} & \gate{R_x(\theta_{12})} & \qw & \qw & \gate{R_y(\theta_{15})} & \ctrl{-3} & \qw
    \end{quantikz}
    \end{subfigure}\\[.2cm]%
    \begin{subfigure}[t]{\textwidth}\centering
        \caption{Scheme b}
        \label{fig:scheme_b}
        \begin{quantikz}[row sep=0.1cm]
            & \gate{R_x(\theta_1)} & \gate{R_z(\theta_2)} & \gate{R_x(\theta_3)} & \ctrl{1}\gategroup[wires=4,steps=4,style={dashed, inner sep=6pt, rounded corners}]{} & \qw & \qw & \gate{R_y(\theta_{13})} & \qw \\
            & \gate{R_x(\theta_4)} & \gate{R_z(\theta_5)} & \gate{R_x(\theta_6)} & \targ{} & \ctrl{1} & \qw & \gate{R_y(\theta_{14})} & \qw \\
            & \gate{R_x(\theta_7)} & \gate{R_z(\theta_8)} & \gate{R_x(\theta_9)} & \qw & \targ{} & \ctrl{1} & \gate{R_y(\theta_{15})} & \qw \\
            & \gate{R_x(\theta_{10})} & \gate{R_z(\theta_{11})} & \gate{R_x(\theta_{12})} & \qw & \qw & \targ{} & \gate{R_y(\theta_{16})} & \qw
    \end{quantikz}
    \end{subfigure}\\[.2cm]%
    \begin{subfigure}[t]{\textwidth}\centering
        \caption{Scheme c}
        \label{fig:scheme_c}
        \vspace{9pt}
         \includegraphics[width=0.85\linewidth]{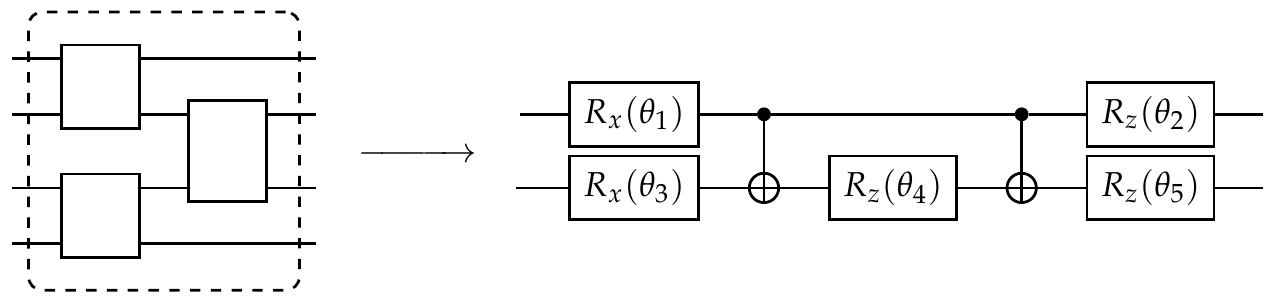}
    \end{subfigure}%
    \caption{ Three schemes
 of ansatzes adopted to do circuit learning for preparing coherent states by using $4$ qubits, where the block (dashed line) in each subplot represents one layer in its corresponding circuit. Such a layer needs to be repeated in the circuit in order to obtain the desired fidelity. \mbox{(\textbf{a}) Hardware} Efficient Ansatz where the rotation angles of all the gates used in the circuit are the variational parameters. (\textbf{b}) Hardware Efficient Ansatz where the rotation angles of $R_y$ gates are set undetermined. (\textbf{c}) Checkerboard Ansatz whose unit of the neighboring two-qubit interaction with variational angles is depicted in detail on the right side.}
    \label{hard}
\end{figure}

Essentially, a core component of a VQA is to find the appropriate ansatz which produces the coherent state with few gates and shallow depth in the circuit.
Here, as shown in Figure \ref{hard}, we apply two kinds of ansatzes by using $4$ qubits: Hardware Efficient Ansatz (Figure \ref{hard}a and~Figure \ref{hard}b, named as Scheme a and Scheme b, respectively) and Checkerboard Ansatz \cite{Checker-board-ansatz} (Figure \ref{hard}c, named as Scheme c), since they are easy to be implemented in the current and near-term hardware. 
 For the
 Hardware Efficient Ansatz which consists of a sequence of single-qubit $R_x$, $R_z$, $R_x$ gates and 
 two-qubit-entangling Controlled-$R_y$ gates, 
  we implemented two alternatives in order to see the relation between the fidelity and the number of control parameters. In Scheme a, the rotation angles of all the gates used in the circuit are the variational parameters with a number of $4N$ in each layer, where $N$ is the number of applied qubits. While the initial values of all the angles are set to $1$ empirically, their optimized values are derived from VQA. For Scheme b, the block of $R_y$ gates, with a number of $N$ (equal to the
  qubit number) labeled in the block as shown in Scheme b,
   is repeated in each layer. 
   In order to gain the desired fidelity, a number $M_l$ of repetition on each layer in the circuit is needed. For a circuit with $M_l$ layers, the number of control parameters in Scheme a and b are $4NM_l$ and $(3+M_l)N$, respectively.
The 
Checkerboard Ansatz, composed by $N-1$ blocks connecting neighboring qubits with a sequence of single-qubit $R_x$ and $R_z$ gates and two-qubit-entangling CNOT gates, is demonstrated in Scheme c, with a total amount of $5(N-1)M_l$ parameters. 

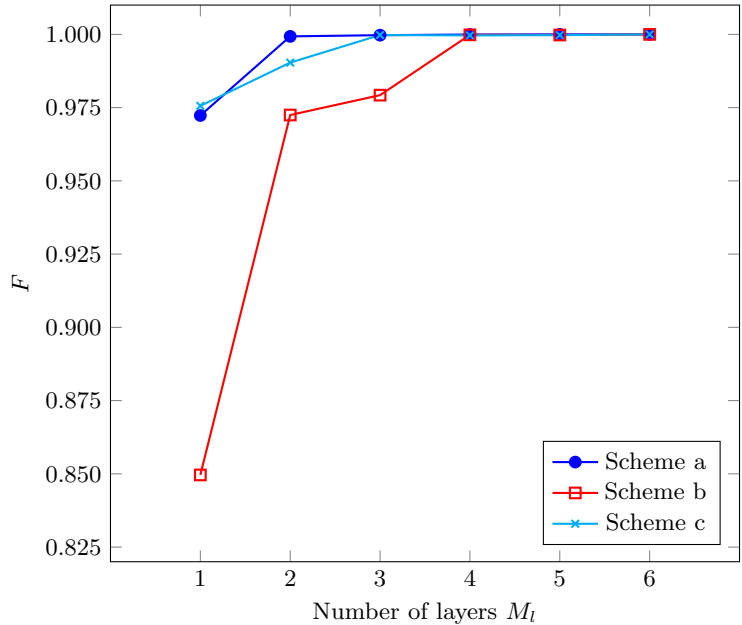
\begin{figure}[!htbp]
    \begin{tikzpicture}
        \begin{axis}[
            height=0.5\textwidth,
            width=0.55\textwidth,
            xtick={1,2,...,6},
            domain = 0:7,
            xlabel={Number of layers $M_l$},
            ylabel={$F$},
            ytick distance=0.025,
            xtick distance=2,
            y tick label style={
                /pgf/number format/.cd,
                fixed,
                fixed zerofill,
                precision=3,
                /tikz/.cd
            },
            ymax=1.01, ymin=0.82,
            xmin=0, xmax=7,
            legend pos=south east,
        ]
        \addplot[blue,thick,mark=*] table[x=depth, y=hea1]{\ansatzfidelitydata};\addlegendentry{Scheme a}
        \addplot[red,thick,mark=square] table[x=depth, y=hea2]{\ansatzfidelitydata};\addlegendentry{Scheme b}
        \addplot[cyan,thick,mark=x] table[x=depth, y=checker]{\ansatzfidelitydata};\addlegendentry{Scheme c}
        \end{axis}
  \end{tikzpicture}
    \caption[ Dependence of fidelity $F$ on the number of layers $M_l$ for preparing the target coherent state $\ket{{1+i}}$ with the application of three schemes whose ansatzes are illustrated in Figure \ref{hard}.]{ Dependence
 of fidelity $F$ on the number of layers $M_l$ for preparing the target coherent state $\ket{{1+i}}$ with the application of the three schemes presented in Figure \ref{hard}.}
    \label{different_qubits}
\end{figure}

The first two columns of 
Table \ref{tab:pagenum}
 shows the number of the applied single-qubit and CNOT gates for three schemes in terms of the number of layers $M_l$ and the number of qubits $N$, where controlled-$R_y$ gates are decomposed into single-qubit and CNOT gates for comparison.  
In order to guarantee high-precision preparation of coherent states ($F>0.9999$), Schemes a, b,
 and c need $4$, $6$,
  and $6$ layers for a 4-qubit system, respectively, as shown in Figure \ref{different_qubits}. 
Scheme b requires less quantum gates and controlled parameters even though it demands more layers.\par

For VQAs, the running time of the optimization is determined by the iteration steps, the number of parameters to be optimized, the initial value of the parameters,
 and the chosen optimizer. The last two columns of the Table \ref{tab:pagenum} show the minimized iteration times and depth to guarantee high-precision preparation of coherent states ($F>0.9999$). The minimal steps needed to realize a high-precision preparation of coherent states $|{1+i}\rangle$ is achieved 
  by Scheme b.

\begin{table}[!htbp]
\setlength{\tabcolsep}{4mm}
 \caption{Quantum gates used for the three schemes shown in~Figure \ref{hard} and their
  corresponding results for acquiring $F>0.9999$.}
 \label{tab:pagenum}
 \begin{tabular}{ccccc}
  \toprule
    \multirow{2}{*}{\vspace{-6pt}\textbf{Scheme}} & \multicolumn{2}{c}{\textbf{Number of Gates}} & \multicolumn{2}{c}{\textbf{Minimized Results for} \boldmath{$F>0.9999$}} \\ \cmidrule{2-5}
    & \multicolumn{1}{c}{\textbf{Single-Qubit}} & \multicolumn{1}{c}{\textbf{CNOT}} & \multicolumn{1}{c}{\textbf{Iteration Times}} & \multicolumn{1}{c}{\textbf{Depth}} \\
  \midrule
   a  & $5NM_l$ & $2NM_l$ & 4166 & 4\\
   b  & $(3+M_l)N$ & $(N-1)M_l$ & 2517 & 6\\
   c  & $5(N-1)M_l$ & $2(N-1)M_l$ & 4099 & 6\\
  \bottomrule
 \end{tabular}
\end{table}

\section{Conclusions}
\label{conclusion}
Coherent states and their digital quantum simulation are of significance in many fields, as a fundamental element in quantum computing, quantum machine leaning, and quantum optics. In this article, we proposed a new method for digitally simulating coherent states in quantum circuits. By expressing the Fock states with an appropriate number of qubits, we decomposed the displacement operator into Pauli matrices via the second quantization. A generalized formula on the Trotter expansion of the displacement operator with $N$ qubits was demonstrated. The derived coherent states in quantum circuits coincided with a Poissonian distribution with high fidelity. Moreover, we also generated coherent states by gradient-based quantum circuit learning. Hardware Efficient Ansatz and Checkerboard Ansatz were used to find coherent states. Different schemes with distinct numbers of variational parameters were compared in terms of quantum resources and iteration times.

{In the NISQ era, seeking efficient encoding methods, i.e., shorter circuit depth and less quantum gates while maintaining high fidelities, is always the object for the digital quantum simulation.  
Finding
 the optimized ansatzes to generate coherent states which can exhibit robustness in the presence of different kinds of noise will also be considered. 
 These will be explored and addressed in our future work. Meanwhile, the transfer of digital quantum simulation from a coherent state to a squeezed state might be interesting to study. The quantum information processing hardware containing continuous-variable objects, such as mechanical or electromagnetic oscillators, instead of discrete-variable qubits, have demonstrated advantages in some aspects, 
  such as quantum error corrections~\cite{error-corrections} and data encoding \cite{encoding-advantages}, although the physical implementation still demands further development~\cite{further-development}. }
We hope our method of digital simulation on coherent states will be useful for the dynamic simulation of quantum many-body systems and device design for quantum machine learning.  

\vspace{6pt}

\appendix
\section{Decomposition of the Displacement Operator}
\label{APPENDIX-A}

{First of all, we start by computing $Z_1=i(\hat{a}^\dagger - \hat{a})$ and $Z_2=-(\hat{a}+\hat{a}^\dagger)$ 
{using \mbox{Equation~\ref{eq:z1_z2}}} with $n=2^N$.}
%
 Observe that $Z_1$ is Hermitian and $Z_2$ is symmetric; so, an odd (even) number of $Y$ matrices, named as $n_y$ must appear in the Pauli string decomposition of $Z_1$ and $Z_2$.  
 Its corresponding Pauli strings can be defined as $(\sigma^D)^{\otimes N-m}(\sigma^S)^{\otimes m}  \text{ }\forall m\in\{1,\dots,N\}$.  For computing all of them,
  we 
   need to iterate over $m$ and compute all of the possible strings taking care about the parity of $n_y$. For a general number of qubits $N$, it can be seen that we  need $N2^{N-1}$ Pauli strings for decomposing each matrix. The weights of the Pauli strings for $Z_1$ are given by

\begin{small}
\begin{equation*}
\frac{(-1)^{\gamma_m}}{2^{N-1}} \sum_{k=0}^{2^{N-m}-1} (-1)^{\epsilon_k}\sqrt{2^{m-1}(2k+1)} \quad \forall m\in\{1,\dots,N\}.
\end{equation*}
\end{small}%
with $\gamma_m=(-1)^{(\lfloor n_y/2\rfloor +  \tilde{n}_y)\text{ mod }2}$, $\tilde{n}_y$ the amount of $Y$ matrices in the $(m-1)$-rightmost $\sigma^S$ matrices of the composition,
 and
\begin{equation}\label{eq:epsilon_sign}
\epsilon_k=\begin{cases}
     0 & \text{ if }k=0\\
     \displaystyle\sum_{j=0}^{\lfloor \log_2k \rfloor} \left(\left\lfloor\frac{k}{2^j}\right\rfloor \text{ mod } 2 \right)i_{j+m}  & \text{ otherwise}
\end{cases}.
\end{equation}

The weights of the Pauli strings for $Z_2$ are given from

\begin{equation*}
\begin{aligned}
\frac{(-1)^{1+\gamma_m}}{2^{N-1}} \sum_{k=0}^{2^{N-m}-1} (-1)^{\epsilon_k}\sqrt{2^{m-1}(2k+1)} \quad \forall m\in\{1,\dots,N\}.
 \end{aligned}
\end{equation*}

We can rearrange these equations to redefine $Z_i$ in terms of its Pauli decomposition, where

\begingroup
\makeatletter\def\f@size{8.8}\check@mathfonts
\def\maketag@@@#1{\hbox{\m@th\normalsize\normalfont#1}}%
\begin{equation*}
 \begin{aligned}
  Z_1 &= \frac{1}{2^{N-1}}\sum_{m=1}^N\sum_{\substack{i_0, \dots, i_{N-1}\\\text{s.t. }n_y\text{ odd}}}\sum_{k=0}^{2^{N-m}-1}(-1)^{\gamma_m+\epsilon_k}\sqrt{2^{m-1}(2k+1)}\left(\sigma_{i_{N-1}}^D\dots\sigma_{i_m}^D\sigma^S_{i_{m-1}}\dots\sigma_{i_0}^S\right) = \sum_{l=1}^{N2^{N-1}}c_lA_1(l), \\
  Z_2 &= \frac{-1}{2^{N-1}}\sum_{m=1}^N\sum_{\substack{i_0, \dots, i_{N-1}\\\text{s.t. }n_y\text{ even}}}\sum_{k=0}^{2^{N-m}-1}(-1)^{\gamma_m+\epsilon_k}\sqrt{2^{m-1}(2k+1)}\left(\sigma_{i_{N-1}}^D\dots\sigma_{i_m}^D\sigma^S_{i_{m-1}}\dots\sigma_{i_0}^S\right) = -\sum_{l=1}^{N2^{N-1}}c_lA_2(l), \\
 \end{aligned}
\end{equation*}
\endgroup
where $c_l$ and $A_{1,2}(l)$ are suitable definitions of each of the $N2^{N-1}$ constants and strings, respectively,
 which this sum returns in order to have a more compact result. 
 Thus,

\begin{small}
\begin{equation*}
 \begin{aligned}
  \hat{D}(\alpha=a+ib) & \approx \left(e^{-iaZ_1/M}e^{-ibZ_2/M}\right)^M 
  \\ &= \left[\left(\prod_{l=1}^{N2^{N-1}}e^{-iac_lA_1(l)/M}\right)\left(\prod_{l=1}^{N2^{N-1}}e^{ibc_lA_2(l)/M}\right)\right]^M
 \end{aligned}
\end{equation*}
\end{small}

Here,
 we compute string-by-string $Z_1$ and $Z_2$ for $n=3$:

\begin{scriptsize}
\begin{equation*}
 \begin{aligned}
  Z_1:& \begin{cases}
         m=1: &
  \begin{array}{rr}\def\arraystretch{3}
    II\sigma_y^2: & \displaystyle\frac{\sqrt{1} + \sqrt{3} + \sqrt{5} + \sqrt{7}}{4} \\[.4em]
    I\sigma_z^1\sigma_y^2: & \displaystyle\frac{\sqrt{1} - \sqrt{3} + \sqrt{5} - \sqrt{7}}{4} \\[.4em]
    \sigma_z^0I\sigma_y^2: & \displaystyle\frac{\sqrt{1} + \sqrt{3} - \sqrt{5} - \sqrt{7}}{4} \\[.4em]
    \sigma_z^0\sigma_z^1\sigma_y^2: & \displaystyle\frac{\sqrt{1} - \sqrt{3} - \sqrt{5} + \sqrt{7}}{4}
   \end{array} \\[5em]
  m=2: &
  \begin{array}{rr}
    I\sigma_x^1\sigma_y^2: & -\displaystyle\frac{\sqrt{2} + \sqrt{6}}{4} \\[.4em]
    I\sigma_y^1\sigma_y^2: & -\displaystyle\frac{\sqrt{2} + \sqrt{6}}{4} \\[.4em]
    \sigma_z^0\sigma_x^1\sigma_x^2: & -\displaystyle\frac{\sqrt{2} - \sqrt{6}}{4} \\[.4em]
    \sigma_z^0\sigma_y^1\sigma_y^2: & -\displaystyle\frac{\sqrt{2} - \sqrt{6}}{4}
   \end{array} \\[5em]
  m=3: &
  \begin{array}{rr}
    \sigma_x^0\sigma_x^1\sigma_y^2: & {-\displaystyle\frac{1}{2}} \phantom{\frac{\displaystyle\sqrt{1}}{4}} \\[.4em]
    \sigma_x^0\sigma_y^1\sigma_x^2: & {-\displaystyle\frac{1}{2}} \phantom{\frac{\displaystyle\sqrt{1}}{4}}  \\[.4em]
    \sigma_y^0\sigma_x^1\sigma_x^2: & {\displaystyle\frac{1}{2}} \phantom{\frac{\displaystyle\sqrt{1}}{4}}  \\[.4em]
    \sigma_y^0\sigma_y^1\sigma_y^2: & {-\displaystyle\frac{1}{2}} \phantom{\frac{\displaystyle\sqrt{1}}{4}}
   \end{array} \\[4.6em]
        \end{cases}\quad&
  Z_2:& \begin{cases}
         m=1: &
  \begin{array}{rr}\def\arraystretch{3}
    II\sigma_y^2: & -\displaystyle\frac{\sqrt{1} + \sqrt{3} + \sqrt{5} + \sqrt{7}}{4} \\[.4em]
    I\sigma_z^1\sigma_y^2: & -\displaystyle\frac{\sqrt{1} - \sqrt{3} + \sqrt{5} - \sqrt{7}}{4} \\[.4em]
    \sigma_z^0I\sigma_y^2: & -\displaystyle\frac{\sqrt{1} + \sqrt{3} - \sqrt{5} - \sqrt{7}}{4} \\[.4em]
    \sigma_z^0\sigma_z^1\sigma_y^2: & -\displaystyle\frac{\sqrt{1} - \sqrt{3} - \sqrt{5} + \sqrt{7}}{4}
   \end{array} \\[5em]
  m=2: &
  \begin{array}{rr}
    I\sigma_x^1\sigma_y^2: & -\displaystyle\frac{\sqrt{2} + \sqrt{6}}{4} \\[.4em]
    I\sigma_y^1\sigma_y^2: & \displaystyle\frac{\sqrt{2} + \sqrt{6}}{4} \\[.4em]
    \sigma_z^0\sigma_x^1\sigma_x^2: & -\displaystyle\frac{\sqrt{2} - \sqrt{6}}{4} \\[.4em]
    \sigma_z^0\sigma_y^1\sigma_y^2: & \displaystyle\frac{\sqrt{2} - \sqrt{6}}{4}
   \end{array} \\[5em]
  m=3: &
  \begin{array}{rr}
    \sigma_x^0\sigma_x^1\sigma_y^2: & {-\displaystyle\frac{1}{2}} \phantom{\frac{\displaystyle\sqrt{1}}{4}} \\[.4em]
    \sigma_x^0\sigma_y^1\sigma_x^2: & {\displaystyle\frac{1}{2}} \phantom{\frac{\displaystyle\sqrt{1}}{4}}  \\[.4em]
    \sigma_y^0\sigma_x^1\sigma_x^2: & {-\displaystyle\frac{1}{2}} \phantom{\frac{\displaystyle\sqrt{1}}{4}}  \\[.4em]
    \sigma_y^0\sigma_y^1\sigma_y^2: & {-\displaystyle\frac{1}{2}} \phantom{\frac{\displaystyle\sqrt{1}}{4}}
   \end{array} \\[4.6em]
        \end{cases}
 \end{aligned}
\end{equation*}
\end{scriptsize}

\section{Accuracy of the Coherent State Expressed in the Truncated Space}
\label{APPENDIX-B}
[-10]{Based on the coherent state in the Fock basis with infinite number of states (\mbox{Equation \ref{target-state}}),} the coherent state in the truncated subspace with $N$ qubits, leading to $2^N$ available independent states, can be expressed as
\begin{equation}\label{eq:poisson_trunc}
 \ket{\psi_\text{tar}'} = \ket{\alpha'} = \frac{e^{-|\alpha|^2/2}\sum_{k=0}^{2^N-1}\frac{\alpha^k}{\sqrt{k!}}\ket{k}}{\sqrt{e^{-|\alpha|^2}\sum_{k=0}^{2^N-1}\frac{|\alpha|^{2k}}{k!}}} 
 = \frac{e^{-|\alpha|^2/2}\sum_{n=0}^{2^N-1}\frac{\alpha^k}{\sqrt{k!}}\ket{k}}{\sqrt{\Gamma(2^N, |\alpha|^2)/\Gamma(2^N)}},
\end{equation}
where $\Gamma(k+1)=k!$ is the gamma function and $\Gamma(k, x)$ is the incomplete gamma function and is defined as

\begin{equation*}
 \Gamma(k, x) = \int_x^\infty t^{k-1}e^{-t}\mathrm{d}t.
\end{equation*}
Thus, the probability of measuring the state $\ket{m}$ is

\begin{equation*}
 P'_m = |\langle m|\psi_\text{tar}'\rangle|^2 = \frac{e^{-|\alpha|^2}\frac{|\alpha|^{2m}}{m!}}{\Gamma(2^N, |\alpha|^2)/\Gamma(2^N)} = \frac{e^{-\langle n\rangle}\frac{\langle n\rangle^m}{m!}}{\Gamma(2^N, \langle n\rangle)/\Gamma(2^N)} = \frac{\Gamma(2^N)}{\Gamma(2^N, \langle n\rangle)}P_m,
\end{equation*}
with $P_m$ the Poissonian probability given by~Equation \ref{probability}. For the particular case of $\ket{\alpha}=|{1+i\rangle}$, we can find $\langle n\rangle = \langle \hat{a}^\dagger\hat{a} \rangle = |\alpha|^2 =2$, since $\Gamma(2^N)/\Gamma(2^N, 2)=21e^2/155\approx 1.0011$ for $N=3$, we can state that $P'_m\approx P_m$.

%


\end{document}